\documentclass[referee]{aa}

\usepackage{graphicx}
\usepackage{color}

\begin{document}

\title{Antideuterons as a probe of primordial\\
black holes}

\author{Aur\'elien Barrau\inst{1,3}, Ga\"elle Boudoul\inst{1,3},
Fiorenza Donato\inst{2,5}, David Maurin\inst{2},\\
Pierre Salati\inst{2,4}, Ilan St\'efanon\inst{1,3},  Richard Taillet\inst{2,4}
}

\offprints{A. Barrau} \mail{barrau@isn.in2p3.fr}

\institute{
ISN Grenoble, 53 av des Martyrs, 38026 Grenoble cedex, France
\and
LAPTH, B.P. 110 Annecy-le-Vieux, 74941, France
\and
Universit\'e Joseph Fourier, Grenoble, 38000, France.
\and
Universit\'e de Savoie, Chamb\'ery, 73011, France.
\and
Universit\`a degli Studi di Torino and INFN, Torino, Italy.
}

\date{Received 19 July 2002; Accepted 17 October 2002}

\titlerunning{Antideuterons from primordial black holes}

\authorrunning{Barrau et al.}

\abstract{
In most cosmological models, primordial black holes ({\sc pbh}) should have 
formed in the early Universe. Their Hawking evaporation into particles could
eventually lead to the formation of antideuterium nuclei.
This paper is devoted to a first computation of this antideuteron flux.
The production of these antinuclei is studied with a simple
coalescence scheme, and their propagation in the Galaxy is treated with
a well-constrained diffusion model.
We compare the resulting primary flux to the secondary background, due
to the spallation of protons on the interstellar matter.
Antideuterons are shown to be a very sensitive probe for 
primordial black
holes in our Galaxy. The next generation of experiments should
allow investigators to significantly improve the current upper limit, nor even
provide the first
evidence of the existence of evaporating black holes.
}

\maketitle

\section{Introduction}

Very small black holes could have formed in the early Universe from
initial density
inhomogeneities (Hawking \cite{Hawking2}), from phase transition (Hawking
\cite{Hawking3}),
from collapse of cosmic strings (Hawking \cite{Hawking4}) or as a
result of a softening of the
equation of state (Canuto \cite{Canuto}). It was also shown
by Choptuik (Choptuik \cite{Choptuik}) and,
more recently, studied in the framework of double inflation (Kim \cite{Kim}),
that {\sc pbh}s could even
have formed by near-critical collapse in the expanding Universe.

The interest in primordial black holes has been revived in the
last years for several reasons.
On the one hand, new experimental data on gamma-rays
(Connaughton \cite{Connaughton})  and cosmic rays (Barrau{\it et al.}
\cite{barrau4})
together with the construction of
neutrino detectors (Bugaev \& Konishchev \cite{Bugaev}), of extremely high-energy
particle observatories (Barrau \cite{Barrau3})
and of gravitational waves interferometers (Nakamura {\it et al.} \cite{Nakamura}) give
interesting investigational means to look for indirect signatures of
{\sc pbh}s. On the other hand, primordial black holes have
been used to derive interesting limits on the scalar fluctuation
spectrum on very small scales, extremely far from the range accessible to CMB
studies (Kim {\it et al.} \cite{Kim2}, Blais {\it et al.} \cite{Po}).
It was also found that {\sc pbh}s are a useful probe of the
early Universe with a varying
gravitational constant (Carr \cite {Carr2}).
Finally, significant progress has been made in the understanding of
the evaporation
mechanism itself, both at usual energies (Parikh \& Wilczek \cite{Parikh}) and in the 
near-Planckian
tail of the spectrum (Barrau \& Alexeyev \cite{Barrau2}, Alexeyev {\it et
al.} \cite{Stas}, Alexeyev {\it et al.} \cite{Stas2}).

For the time being there is no evidence in experimental data in favour of the
existence of {\sc pbh}s in our Universe. Only upper limits on their 
number density
or on their explosion rate have been obtained  (Barrau {\it et al.}
\cite{barrau4}, MacGibbon \& Carr \cite{MacGibbon2}). As the spectra of gamma-rays,
antiprotons and positrons can be well explained without any new 
physics input ({\it
e.g.} {\sc pbh}s or annihilating supersymmetric particles) there is 
no real hope
for any detection in the forthcoming years using those cosmic-rays. The 
situation is
very different with antideuterons which could be a powerful probe used to search for
exotic objects, as the
background is extremely low below a few GeV (Chardonnet {\it et al.}
\cite{chardonnet}, Donato {\it et al.} \cite{fiorenza}).
Such light antinuclei could be the only way to find {\sc pbh}s or to improve the
current limits. This paper is organized along the same guidelines as our
previous study on {\sc pbh} antiprotons (Barrau {\it et al.} \cite{barrau4}), to which the reader is
referred for a full description of the source and propagation model used.
The main difference is the necessity to consider a coalescence scheme 
for the antideuteron production. 
We compute the expected flux of antideuterons
for a given distribution of {\sc pbh}s in our Galaxy, propagate
the resulting spectra in a
refined astrophysical model whose parameters are strongly constrained and,
finally, give
the possible experimental detection opportunities with the next generation of
experiments as a function of the uncertainties on the model.

\section{Antideuterons emission}

\subsection{Hawking process and subsequent fragmentation}

The Hawking black hole evaporation process can be intuitively
understood as a quantum creation of
particles from the vacuum by an external field. The basic
characteristics can be easily seen
through a simplified model (see Frolov \& Novikov (\cite{Frolov}) for
more details) which allowed Schwinger to derive, in 1951, the rate of
particle production by a uniform electric field
and remains correct, at the intuitive level, for black hole
evaporation. If
we focus on a static gravitational field, it should be taken into account
that the energy of a particle
can be written as $E=-p_\mu \xi^\mu$, where $p^\mu$ is the four-momentum
and $\xi^\mu$ is the Killing
vector. The momentum being a future-directed timelike vector, the
energy $E$ is always positive in
the regions where the Killing vector is also future-directed and
timelike. If both particles were
created in such a region, their total energy would not vanish and the
process would, therefore, be
forbidden by conservation of energy. As a result, a static
gravitational field can create
particles only in a region where the Killing vector is spacelike.
Such a region lies inside the
Killing horizon, {\it i.e.} the $\xi^2=0$ surface, which is the event
horizon in a static spacetime.
This basic argument shows that particle creation by a gravitational
field in a static spacetime
(this is also true in a stationary case) is possible only if it
contains a black hole. Although
very similar to the effect of particle creation by an electric field,
the Hawking process has a
fundamental difference: since the states of negative energy are
confined inside the hole, only one
of the created particles can appear outside and reach infinity.

The accurate emission process, which mimics a Planck law,
was derived by Hawking, using the usual quantum
mechanical wave equation for a collapsing object with a postcollapse
classical curved metric instead
of a precollapse Minkowski one (Hawking \cite{Hawking5}). He found
that the emission spectrum for
particles of energy $Q$ per unit of time $t$ is, for each degree of
freedom:
\begin{equation}
\frac{{\rm d}^2N}{{\rm d}Q{\rm
d}t}=\frac{\Gamma_s}{h\left(\exp\left(\frac{Q}{h\kappa/4\pi^2c}\right)-(-1)^{2s}\right)}
\end{equation}

where contributions of angular velocity and electric potential have
been neglected since the
black hole discharges and finishes its rotation much faster than it
evaporates (MacGibbon \& Webber \cite{Gibbons}, Page \cite{Page1}). $\kappa$ is the surface 
gravity, $s$ is the
spin of the emitted species and
$\Gamma_s$ is the absorption probability. If we introduce the Hawking
temperature (one of the rare
physical formul\ae~using all the fundamental constants) defined by
\begin{equation}
T=\frac{hc^3}{16\pi k G M}\approx\frac{10^{13}{\rm g}}{M}{\rm GeV}
\end{equation}
the argument of the exponent becomes simply a function of $Q/kT$.
Although the absorption probability is often approximated by its
relativistic limit
\begin{equation}
\lim_{Q \rightarrow \infty} \Gamma_s =
\frac{108\pi^2G^2M^2Q^2}{h^2c^6}
\end{equation}
we took into account in this work its real expression for
non-relativistic particles:
\begin{equation}
\Gamma_s=\frac{4\pi \sigma_s(Q,M,\mu)}{h^2c^2}(Q^2-\mu^2)
\end{equation}
where $\sigma_s$ is the absorption cross-section computed numerically
(Page \cite{Page2}) and $\mu$ the rest mass of the emitted particle.
Even if this mass effect is partially
compensated by the pseudo-oscillating behaviour of the cross-section
  and remains at the level
of a correction, we found some substantial discrepancies between the
geometric limit and the
numerical calculation which justifies this technical complication.\\

As was shown by MacGibbon and Webber (MacGibbon \& Webber \cite{MacGibbon1}), when the
black hole temperature is
greater than the quantum chromodynamics confinement scale
$\Lambda_{QCD}$, quark and gluon jets are
emitted instead of composite hadrons. To evaluate the number of
emitted antinucleons $\bar{N}$ , one therefore
needs to perform the following convolution:
\begin{equation}
\frac{{\rm d}^2N_{\bar{N}}}{{\rm d}E{\rm d}t}=
\sum_j\int_{Q=E}^{\infty}\alpha_j\frac{\Gamma_{s_j}(Q,T)}{h}
\left(e^{\frac{Q}{kT}}-(-1)^{2s_j}\right)^{-1}
\times\frac{{\rm d}g_{j\bar{N}}(Q,E)}{{\rm d}E}{\rm
d}Q
\end{equation}
where $\alpha_j$ is the number of degrees of freedom, $E$ is the
antinucleon energy and
${\rm d}g_{j\bar{N}}(Q,E)/{\rm d}E$ is the normalized differential
fragmentation function, {\it i.e.}
the number of antinucleons between $E$ and $E+{\rm d}E$ created by a
parton jet of type $j$ and energy
$Q$. The fragmentation functions have been evaluated with the
high-energy physics frequently-used event generator
PYTHIA/JETSET (Tj\"{o}strand \cite{Tj}). This program is based on the
so-called string fragmentation
model (developed by the Lund group) which is an explicit and detailed
framework where the long-range
confinement forces are allowed to distribute the energies and
flavours of a parton configuration among a
collection of primary hadrons. It has received many improvements
related, {\it e.g.}, to parton showers,
hard processes, Higgs mechanisms and it is now in excellent agreement
with experimental data.

\subsection{Coalescence scheme}

\begin{figure}
\centerline{\includegraphics*[width=\textwidth]{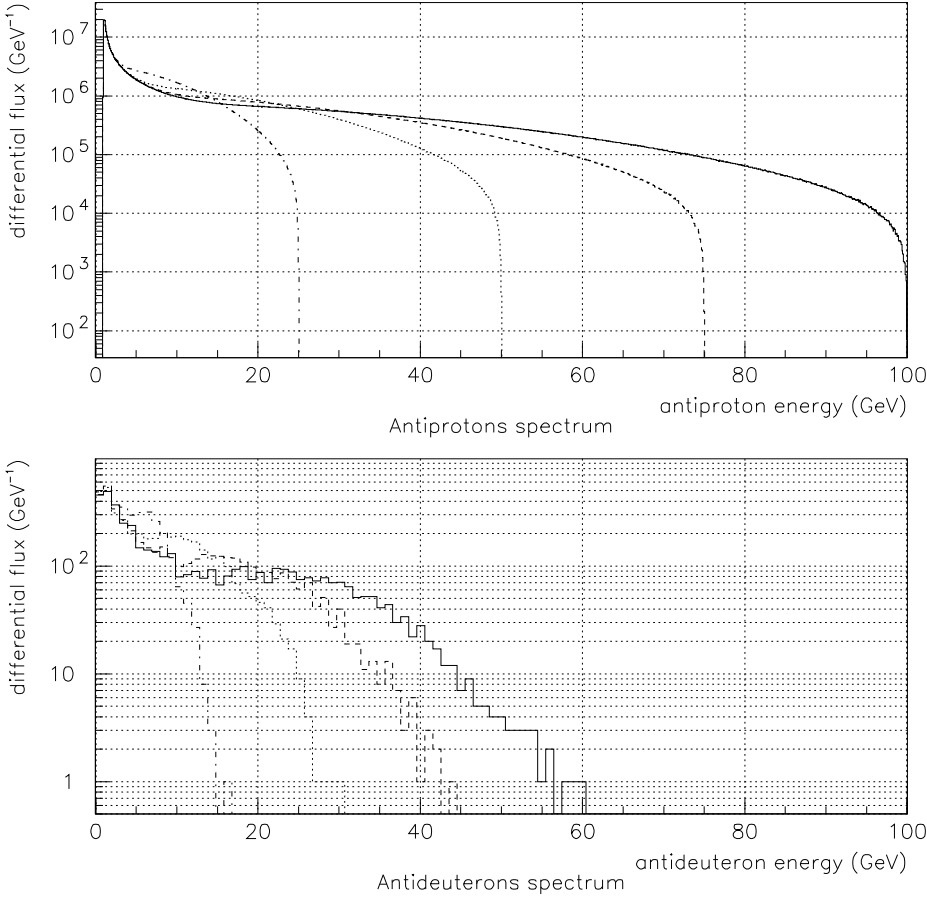}}
\caption{ Upper plot: antiproton differential spectrum obtained with 
$1.9\times 10^8~\bar{u}$ quark jets generated by PYTHIA at 100, 75, 50, 25 GeV.
Lower plot: antideuteron spectrum obtained in the same conditions with a
coalescence momentum $p_0=160$ MeV.}
\label{fig:dbar}
\end{figure}

In the context of proton-nucleus collisions it was suggested that, 
independently
of the details of the deuteron formation mechanism, the momentum 
distribution of
deuterons should be proportional to the product of the proton and neutron
momentum distributions (see Csernai \& Kapusta (\cite{csernai}) for a review). This was based on
phase space considerations alone: the deuteron density in momentum space is
proportional to the product of the proton density and the probability 
of finding a
neutron within a small sphere of radius $p_0$ around the proton momentum. Thus:
\begin{equation}
\gamma\frac{d^3N_d}{dk^3_d}=\frac{4\pi}{3}p_0^3\left(\gamma
\frac{d^3N_p}{dk^3_p}\right)\left(\gamma \frac{d^3N_n}{dk^3_n}\right)
\end{equation}
where $p_0$ is the coalescence momentum which must be determined from
experiments. The very same arguments can be used for antideuterons resulting
from an antiproton and antineutron momentum distribution. In our case, the
coalescence scheme has to be implemented directly within the {\sc pbh} jets as no
nuclear collision is involved. We defined the following procedure:\\
- for each hadronic jet resulting from a parton emitted by a {\sc pbh}, we search for
antiprotons\\
- if an antiproton is found within the jet, we search for antineutrons\\
- if an antineutron is also found within the same jet, we compare their
momenta\\
- if the difference of the antiproton and antineutron momenta is smaller than the coalescence momentum
$p_0$, we consider that an antideuteron should be formed.\\
As the coalescence momentum $p_0$ is not Lorentz
invariant, the condition must be implemented in the correct frame, 
namely in the
antiproton-antineutron center of mass frame instead of the laboratory one. Fig.
\ref{fig:dbar} gives the differential spectrum of antiprotons resulting from 
$1.9\times 10^8~\bar{u}$ quark jets generated at 100, 75, 50, 25 GeV and the
subsequent distribution of antideuterons with $p_0=160$ MeV. The ratio
between the antideuteron and antiproton spectra is of the order of a few times
$10^{-5}$, which reflects the mean amplitude of the cosmic antideuteron
flux from {\sc pbh}s, given in Section 4 of this article, with respect to the one
given for antiprotons in Barrau {\it et al.} (\cite{barrau4}). This value is not
surprising at it is in reasonable agreement with:\\
- the Serpukhov experimental ratio of the
$\bar{p}$ to $\bar{D}$ production cross-sections for proton-proton interactions 
measured at $\sqrt{s}=11.5$ GeV (between $1.9\times 10^{-5}$ and $3.5\times
10^{-5}$ depending on the transverse momentum) as given in Abranov {\it et al.} 
(\cite{abranov}) and for proton-aluminium interactions 
measured at the same energy ($7\times 10^{-5}$ for a center of mass 
transverse momentum around 270 MeV) as given in Binon {\it et al.} (\cite{binon})\\
- the CERN-ISR experimental ratio of the
$\bar{p}$ to $\bar{D}$ production cross-sections for proton-proton interactions
measured at $\sqrt{s}=53$ GeV (between $10^{-4}$ and $3.4\times 10^{-4}$) as given
in Alper {\it et al.} (\cite{alper}) and Gibson {\it et al.} (\cite{gibson})\\
- the theoretical cosmic secondary $\bar{p}$ to $\bar{D}$ ratio (around $10^{-5}$) as given in
Chardonnet {\it et al.} (\cite{chardonnet})\\
- the theoretical cosmic primary $\bar{p}$ to $\bar{D}$  ratio from
neutralinos (around a few times $10^{-4}$) as given in Donato {\it et al.} (
\cite{fiorenza})\\
- the simulated ratio of $\bar{p}$ to $\bar{D}$ fluxes created within the
Earth's
atmosphere (around $10^{-5}$) as evaluated with the program (Derome {\it et al.} 
(\cite{derome})) that was used to
explain AMS experimental data (Derome, private communication).\\

Although the orders of magnitude are correct, large discrepancies between these
theoretical and experimental results can be noticed. This is taken into account in
this work by allowing the coalescence momentum to vary between 60
MeV and 285 MeV, numbers than can be considered as "extreme" possible
values.\\

The flux of emitted antideuterons should now be written as:

\begin{equation}
\frac{{\rm d}^2N_{\bar{D}}}{{\rm d}E{\rm d}t}=
\sum_j\int_{Q=E}^{\infty}\alpha_j\frac{\Gamma_{s_j}(Q,T)}{h}
\left(e^{\frac{Q}{kT}}-(-1)^{2s_j}\right)^{-1}
\times\frac{{\rm d}g_{j\bar{D}}(Q,E,p_0)}{{\rm d}E}{\rm
d}Q
\end{equation}
where ${\rm d}g_{j\bar{D}}(Q,E,p_0)/{\rm d}E$ is the fragmentation 
function into
antideuterons evaluated with this coalescence model for a given
momentum $p_0$. 
As the mean number of produced antideuterons per jet is
extremely low, millions of events were generated for each
energy and each partonic degree of freedom. Some interpolations are 
also required
to avoid a diverging computing time: the associated uncertainties have been
found to be negligible. 

\subsection{Convolution with the mass spectrum}

The above expression gives the antideuteron flux due to a
single black hole of temperature $T$.
As {\sc pbh}s of different temperatures (or masses) should be present,
this flux must be
integrated over the full mass spectrum of {\sc pbh}s:
\begin{displaymath}
       q^{prim}(r,z,E) = \int\frac{d^2N_{\bar{D}}(M,E)}{dE\, dt} \cdot
       \frac{d^2n(r,z)}{dM \, dV} dM
\end{displaymath}
with
\begin{displaymath}
       \frac{{\rm d}n}{{\rm d}M}\propto M^2{\rm~for~M}<M_*
\end{displaymath}
\begin{displaymath}
       \frac{{\rm d}n}{{\rm d}M}\propto M^{-5/2}{\rm~for~M}>M_*
\end{displaymath}
where $M_* \approx 5\times 10^{14} {\rm g}$ is the initial mass of a
{\sc pbh} expiring nowadays. As explained in (Barrau {\it et al.} \cite{barrau4}),
the shape below $M_*$ does not depend on any assumption about the initial mass
spectrum whereas the shape above $M_*$ relies on the assumption of a scale-invariant power spectrum. The resulting distribution is, then,
normalized to the local {\sc pbh} density $\rho_\odot$. 
The spatial dependence of this source term is given in eq. (\ref{halo}).

\section{Propagation and source distribution}
The propagation of the antideuterons produced by {\sc pbh}s in the Galaxy
has been studied in the two zone diffusion model described in Donato {\it et al.} 
(\cite{fio}).\\
In this model, the geometry of the Milky Way is a cylindrical box whose
radial extension is $R=20$
kpc from the galactic center, with a disk whose thickness is $2h=200$ pc and a
diffusion halo  whose extent is still subject to 
large uncertainties.\\
The five parameters used in this model are: $K_0$,
$\delta$ (describing the diffusion coefficient $K(E)=K_0 \beta 
R^{\delta}$), the
halo half height L, the convective velocity $V_c$ and the Alfv\'en velocity
$V_a$. They are varied within a given range determined by an exhaustive and
systematic  study of cosmic
ray nuclei data (Maurin {\it et al.} \cite{david}, \cite{david2}). The same parameters as employed to
study the antiproton flux (Barrau {\it et al.} \cite{barrau4}) are used again in this analysis.\\
The antideuterons density produced by evaporating {\sc pbh}s per 
energy bin $\psi_{\bar
D}$ obeys the following diffusion equation:
\begin{equation}
\left\{V_c\frac{\partial}{\partial z} -K\left(\frac{\partial^2}{{\partial
z}^2}\left(r\frac{\partial}{\partial z}\right)\right)\right\}\psi_{\bar
D}(r,z,E) +2h\delta (z) \Gamma_{\bar D}\psi_{\bar D}(r,0,E)= 
q^{prim}(r,z,E)
\label{eqdiff}
\end{equation}
where $q^{prim}(r,z,E)$ corresponds to the source term discussed at the end of this section.\\
The total collision rate is given by
$\Gamma_{\bar D} = n_H \sigma_{\bar D H}v_{\bar D}$ where 
$\sigma_{\bar D H}$ is
the total antideuteron cross-section with protons (Hagiwara {\it et al.}
\cite{secteff}).
The hydrogen density, assumed to be constant all over the disk, has 
been fixed to
$n_H=1$ cm$^{-3}$.

Performing Bessel transforms, all the quantities can be expanded over the
orthogonal set of Bessel functions of zeroth order:
\begin{equation}
\psi_{\bar D} = \sum_{i=1}^{\infty}N_i^{\bar D,prim}
J_0(\zeta_i (x))
\end{equation}
and the
solution of the equation (\ref{eqdiff}) for antideuterons can be written as:
\begin{equation}
N_i^{\bar D,prim}(0)=\exp\left(\frac{-V_c
L}{2K}\right)\frac{y_i(L)}{A_i\sinh\left(S_iL/2\right)}
\end{equation}
where
\begin{displaymath}
\left\{
\begin{array}{l}
y_i(L) = 2\int_0^L \exp\left(\frac{V_c}{2K}(L-z')\right)\sinh\left(\frac{S_i}{2}(L-z')\right)q_i^{prim}(z')
dz'\\
S_{i} \equiv \left\{
{\displaystyle \frac{V_{c}^{2}}{K^{2}}} \, + \,
4 {\displaystyle \frac{\zeta_{i}^{2}}{R^{2}}}
\right\}^{1/2}\\
A_{i} \equiv 2 \, h \, \Gamma^{ine}_{\bar{D}}
\; + \; V_{c} \; + \; K \, S_{i} \,
{\rm coth} \left\{ {\displaystyle \frac{S_{i} L}{2}} \right\}
\;\;
\end{array}
\right.
\end{displaymath}

Energy changes (predominantly ionization losses, 
adiabatic losses and diffusive reacceleration) are taken into 
account via a second order differential equation for $N_i^{\bar{D},prim}$ 
(see, {\it e.g.} Eq.~(9) in Barrau {\it et al.} (\cite{barrau4}), or Secs.3.6.1, 
3.6.2 and 3.6.3 in Maurin {\it et al.} (\cite{david}) for further details). 
At variance with antiproton studies, in a first approximation, 
we discarded the so-called tertiary term (corresponding to 
nonannihilating inelastic reaction, as given in Sec.~4 from Donato {\it et al.} 
(\cite{fiorenza}) 
which should be unimportant at the considered energies since the binding 
energy of this nucleus is of about 2~MeV.\\

The spatial distribution of {\sc pbh} is a priori unknown.
However, as these objects should have formed in the very early stages of
the history of the Universe, it seems reasonable to assume that their 
distribution should be rather homogeneous.
When the cosmic structures have formed, 
they should have followed the cold 
dark matter particles and we assume that they currently have the same
distribution. As a consequence,
the following profile for the {\sc pbh}s distribution has been used 
(normalized to
the local density):
\begin{equation}
f(r,z)=\frac{R_c^2 + R_\odot^2}{R_c^2 +  r^2 + z^2}
\label{halo}
\end{equation}
where the core radius $R_c$ has been fixed to 3.5 kpc and $R_\odot$=8 kpc.
This profile corresponds to the isothermal case with a spherical symmetry,
the uncertainties on $R_c$ and the consequences of a possible flatness
have been shown to be irrelevant in (Barrau {\it et al.} \cite{barrau4}).

\section{Top of the atmosphere spectrum and experimental detection}

\begin{figure}
\centerline{\includegraphics*[width=\textwidth]{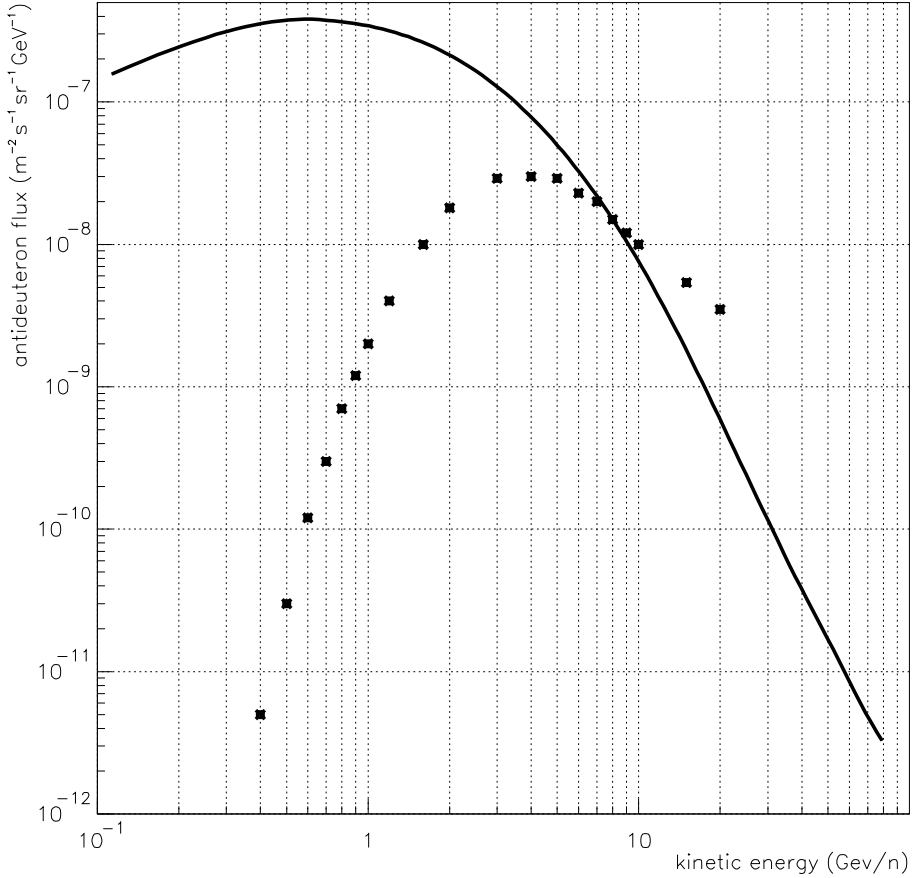}}
\caption{TOA antideuteron flux at solar minimum. The upper curve 
(left part) is from
a {\sc pbh} distribution with a local density of $10^{-33}$ g.cm$^{-3}$
and the lower curve (taken from Donato {\it et al.} \cite{fiorenza}) is 
from secondary processes.}
\label{fig:TOA}
\end{figure}

The flux is then solar modulated in the force field approximation with $\Phi=500$ MV - corresponding to the solar minimum - 
and shown on
Fig. \ref{fig:TOA} for a reasonable ($p_0$=160 MeV/c, $L$=3 kpc) set of parameters at the top of  atmosphere (TOA).
The lower
curve is the antideuteron background due to interactions of cosmic rays on the
interstellar medium as given in (Donato {\it et al.} \cite{fiorenza}) whereas the upper 
curve is due to
evaporating {\sc pbh}s with a local density of $10^{-33}$g.cm$^{-3}$ 
(allowed by
the currently available upper limits (Barrau {\it et al.} \cite{barrau4})). 
Secondaries have been obtained 
in a two-zone diffusion model, with some simplifications: no convection 
and no energy losses have been included.
However, as in the case of antiprotons, their effect should 
be marginal, while they are very important for primary fluxes, and 
the conclusions of the present analysis should not be substantially modified.
To see all the computation steps,
we refer the interested reader to Donato {\it et al.}
(\cite{fiorenza}): the procedure is basically the same as in this
work, except for the production cross-sections that are simply deduced from
the antiproton production cross-sections within a coalescence model with a fixed
momentum (taken as 58 MeV, which corresponds to 116 MeV in our notation)
instead of being computed by a Monte-Carlo method.
The fundamental point
is that this background becomes extremely small below a few GeV/n for 
kinematical
reasons: the threshold for an antideuteron production is $E=17\, 
m_p$ (total energy)
in the laboratory, 2.4 times higher than for antiproton production.
The center of mass is, therefore, moving fast and it is very unlikely 
to produce
an antideuteron at rest in the laboratory.
It should be noted that the secondary  antideuteron background
is only presented here to give a crude estimate of the expected
"physical" background.
In a forthcoming paper, we expect to study this
secondary flux in much more detail, taking special care in the treatment of
diffusion and the cross-sections.
\\

The number of events expected in the AMS experiment (Barrau \cite{Barrau5}) 
onboard the
International Space Station can be estimated, following Donato {\it et al.} 
(\cite{fiorenza}).
Taking into account the geomagnetic rigidity cut-off below
which the cosmic-ray flux is suppressed (as a function of the orbit parameters
), the acceptance of the detector and convoluting with the TOA spectrum, we 
obtain 7
events in three years between 500 MeV/n and 2.7 GeV/n for the previously-given {\sc 
pbh} density and the previously-given typical astrophysical and nuclear parameters.
This is a quite low value which would be difficult to measure due to the possible
mis-reconstruction of $\bar{p}$ or $D$ events. Nevertheless, it 
should  be emphasized that the situation is very different to that of antiprotons, as the 
limit here is not due to the unavoidable physical background but just to the instrument 
capability. Many uncertainties are still unremoved and can affect the primary flux 
more significantly than the secondary one.
\\

\begin{figure}
\centerline{\includegraphics*[width=\textwidth]{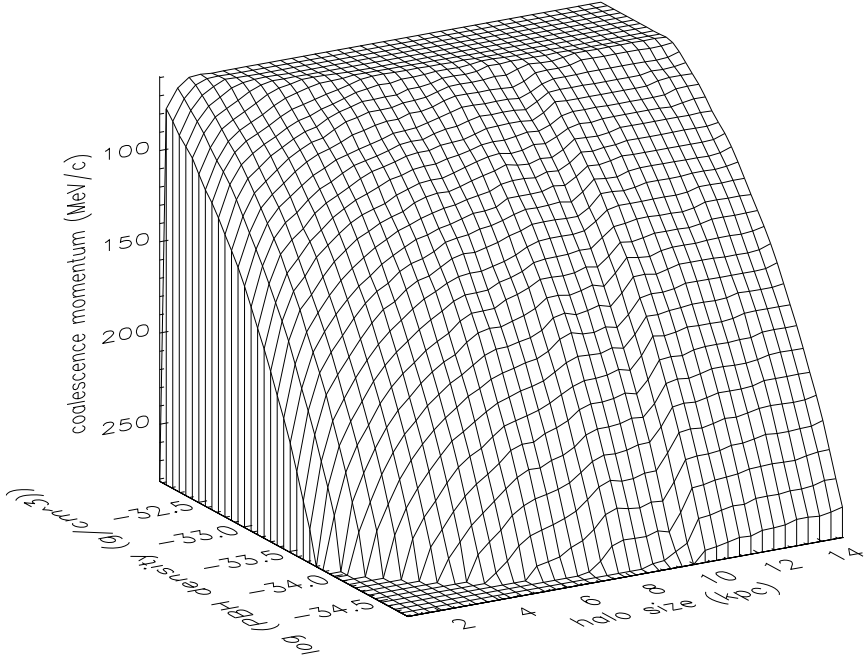}}
\caption{Parameter space (halo thickness $L$: 1-15 kpc; coalescence momentum
$p_0$: 60-285 MeV/c; {\sc pbh} density $\rho_{\odot}$:
$10^{-35}-10^{-31}$g.cm$^{-3}$)
within the AMS sensitivity (3 years of data). The allowed region lies
below the surface.}
\label{fig:3d_ams}
\end{figure}

\begin{figure}
\centerline{\includegraphics*[width=\textwidth]{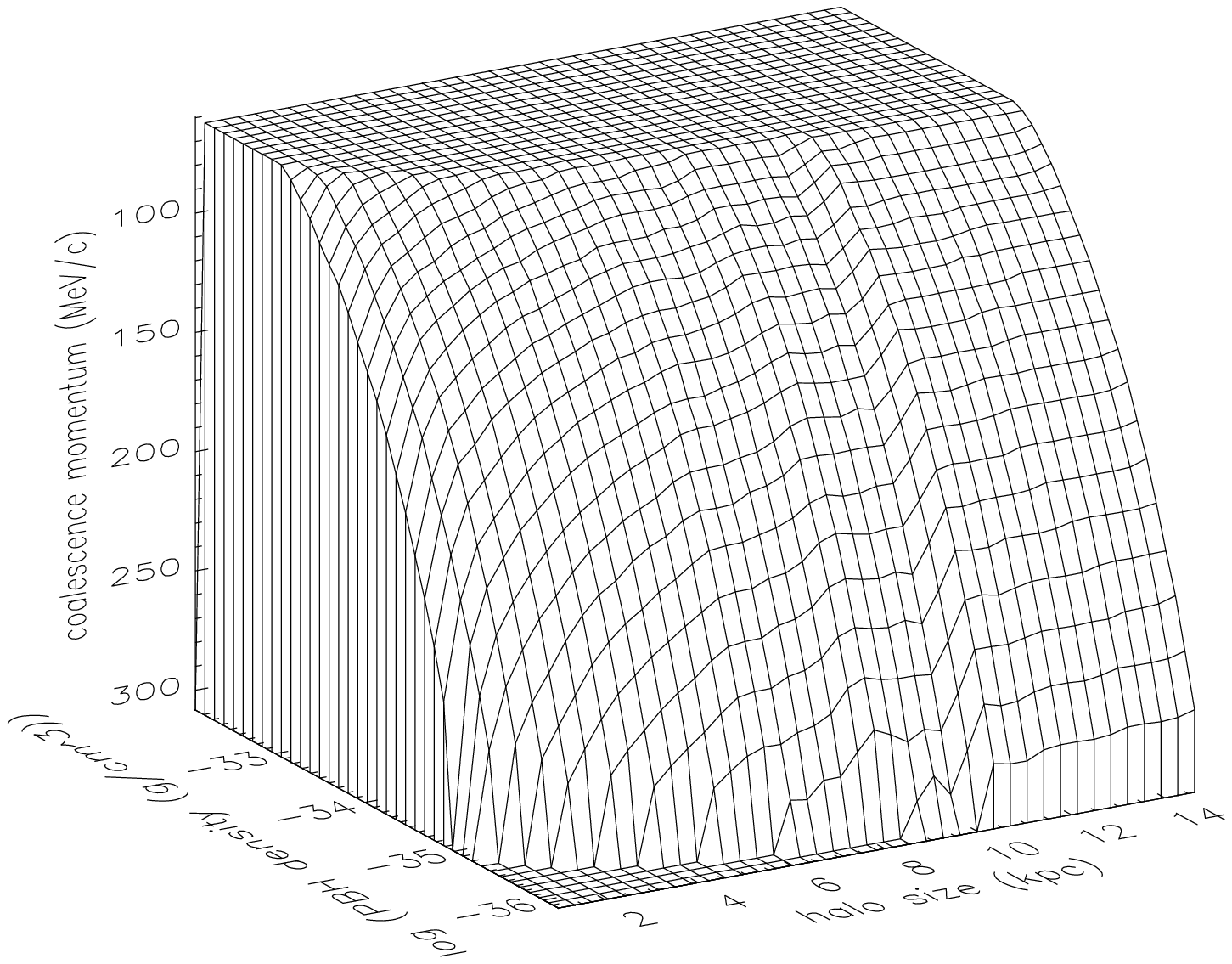}}
\caption{Parameter space (halo thickness $L$: 1-15 kpc; coalescence momentum
$p_0$: 60-285 MeV/c; {\sc pbh} density $\rho_{\odot}$:
$10^{-35}-10^{-31}$g.cm$^{-3}$)
within the GAPS sensitivity (3 years of data taking). The allowed region lies
below the surface.}
\label{fig:3d_gaps}
\end{figure}

\begin{figure}
\centerline{\includegraphics*[width=\textwidth]{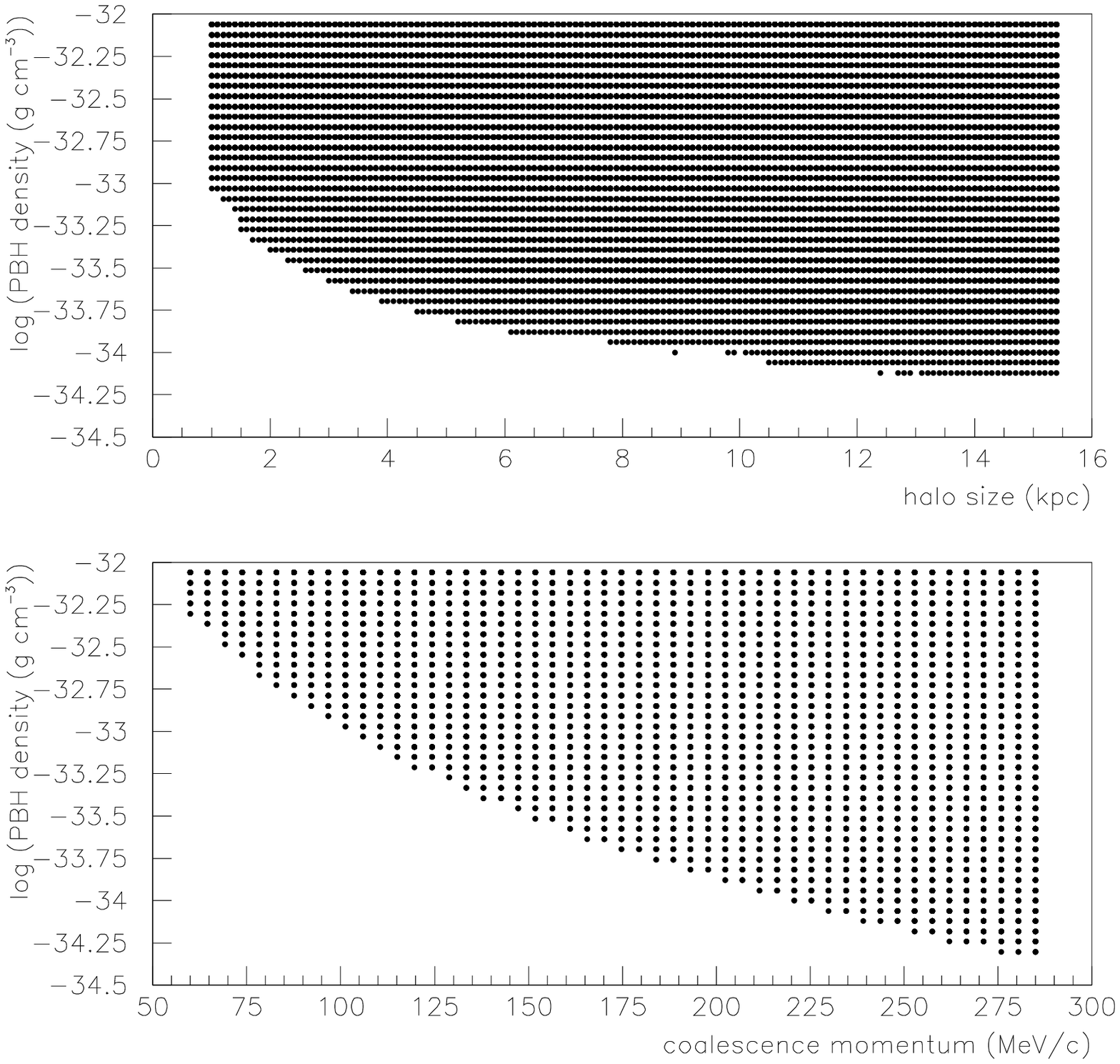}}
\caption{Upper plot: parameter space ({\sc pbh} density vs halo 
thickness) within the
AMS sensitivity for a
fixed value of the coalescence momentum $p_0=160$ MeV/c. Lower plot:
parameter space ({\sc pbh} density vs coalescence momentum) for a 
fixed value of the
halo thickness $L=3$ kpc.}
\label{fig:ams}
\end{figure}

\begin{figure}
\centerline{\includegraphics*[width=\textwidth]{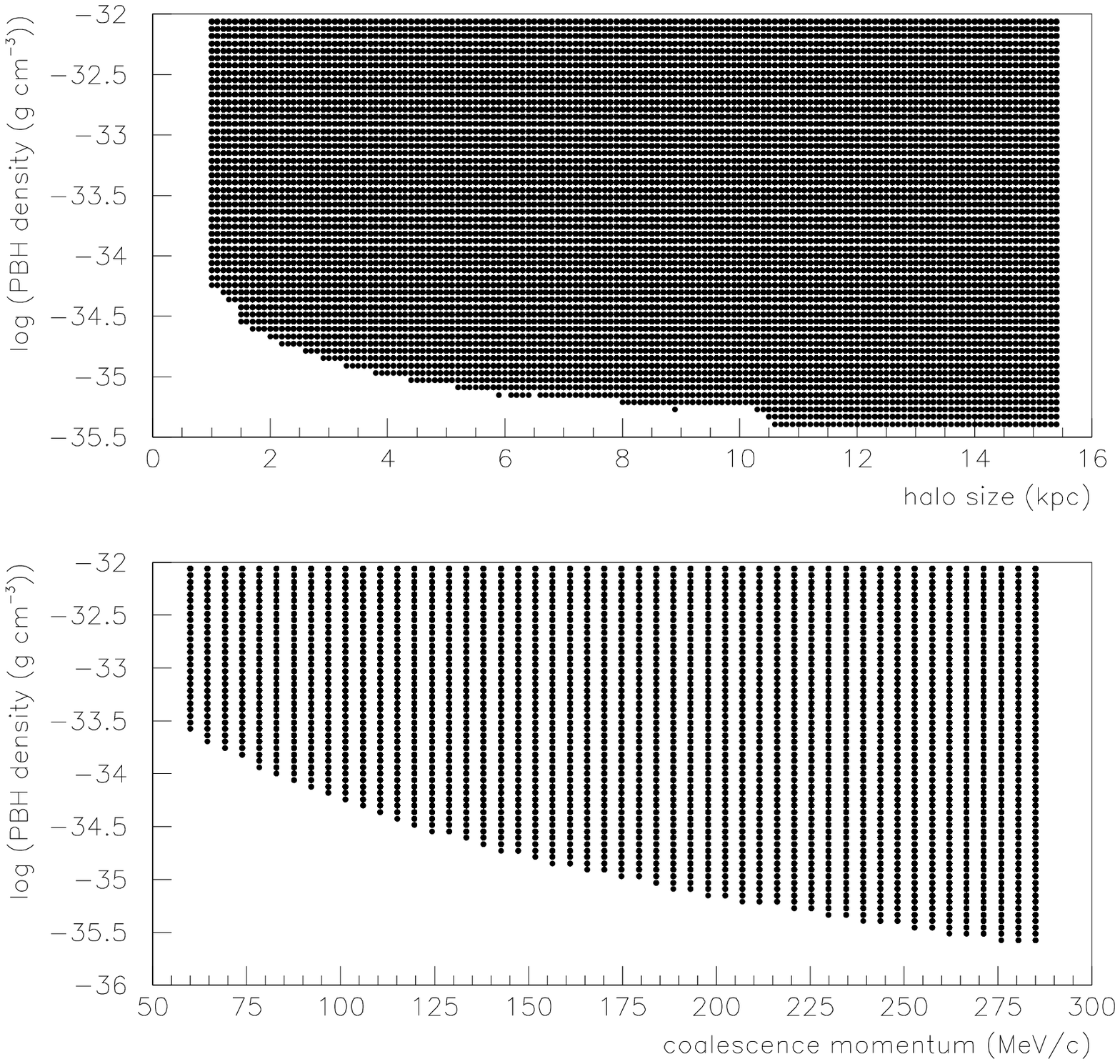}}
\caption{Upper plot: parameter space ({\sc pbh} density vs halo 
thickness) within the
GAPS sensitivity for a
fixed value of the coalescence momentum $p_0=160$ MeV/c. Lower plot:
parameter space ({\sc pbh} density vs coalescence momentum) for a 
fixed value of the
halo thickness $L=3$ kpc.}
\label{fig:gaps}
\end{figure}

In order to be more quantitative, we performed a multi-variable
analysis.
Our model has a large set of free parameters: the astrophysical quantities used for
propagation ($K_0,\delta,L,V_c,V_a$), the local density $\rho_\odot$ 
of {\sc pbh}s
and the
nuclear coalescence momentum $p_0$ for the formation of 
antideuterons. To evaluate
the possible detection of a signal we chose the following strategy: 
as the main
uncertainty from astrophysical processes comes from the halo thickness $L$, the
other parameters were fixed to the value  giving the
smallest flux. This sub-set of parameters depends slightly on $L$
and was varied as a function of $L$ to ensure that whatever the thickness chosen
the real minimum is reached.
All the results are therefore conservative. The 
remaining variables
$\rho_\odot$, $L$ and $p_0$ are then varied within their allowed 
physical ranges: $L$
between 1 and 15 kpc (see Barrau {\it et al.} (\cite{barrau4}) for the details), $p_0$ 
between 60 and 280
MeV/c (depending on the experiments) and $\rho_\odot$ on the largest 
scale matching the related experimental sensitivity.
Two experiments were investigated: the large spectrometer AMS (Barrau
\cite{Barrau5})
which
will take data over 3 years from 2005 and the GAPS project 
(Mori {\it et al.} \cite{GAPS}), based on a
clever design using X-ray desexcitation of exotic atoms. The allowed 
parameter space is
given in Fig.\ref{fig:3d_ams} and Fig.\ref{fig:3d_gaps}: the
values of $L$, $p_0$ and $\rho_\odot$ that can be explored by the considered
experiment, without taking into account possible mis-reconstructions, are located
below the surface. The sensitivity of AMS was taken to be 
$5.7\times 10^{-8}$ m$^{-2}$sr$^{-1}$GeV/n$^{-1}$sec$^{-1}$ between 
500 MeV/n and 2.7 GeV/n for three years of observations whereas the one of GAPS was 
chosen as $2.6\times 10^{-9}$ m$^{-2}$sr$^{-1}$GeV/n$^{-1}$sec$^{-1}$ 
between 0.1 GeV/n and 0.4 GeV/n for the same duration (Mori {\it et al.}
\cite{GAPS}).
To make the results easier
to read, Fig.\ref{fig:ams} and Fig.\ref{fig:gaps} give the accessible 
densities of {\sc
pbh}s for AMS and GAPS with a fixed $L$ (at the more reasonable value 
around 3 kpc)
or a fixed $p_0$ (at the more favoured 
value around 160 MeV/c). 
As expected, the primary flux is increasing  
linearly with the
PBH density (at variance with the search for supersymmetric particles related to the
square of $\rho_{\odot}$, as a collision is involved), linearly with the magnetic
halo thickness (as the core radius $R_c$ is of the same order as $L$) and with
the third power of the coalescence momentum (as the probability to create an
antideuteron is related to a volume element in this space).
The smallest detectable density
of {\sc pbh}s for the employed astrophysical and nuclear parameters is
$\rho_\odot\approx 10^{-33.60}\approx 2.6\times 10^{-34}$ g.cm$^{-3}$ for AMS and 
$\rho_\odot \approx 10^{-34.86}\approx 1.4\times 10^{-35}$ 
g.cm$^{-3}$ for GAPS.
It is much less than the best upper limit available nowadays $\rho_\odot<1.7\times 
10^{-33}$ g.cm$^{-3}$ and
it should open an interesting window for discovery in the forthcoming years. If
no antideuteron is found,
the upper limits will be significantly decreased, allowing stringent 
constraints on
the spectrum of fluctuations in the Universe on very small scales. It 
should also
be mentioned that, in spite of its much smaller acceptance, the 
PAMELA experiment (Adriani {\it et al.} \cite{pam})
could supply interesting additional information thanks to its very low energy
threshold, around 50 MeV/n.

\section{Discussion}

As recently pointed out in Donato {\it et al.} (\cite{fiorenza}),
antideuterons seem to be a more promising probe to look for exotic
sources than antiprotons.
In this preliminary study, we show that this should also be the case for
{\sc pbh}s, so that
antideuterons may be the only probe to look for  such objects.
They should allow a great improvement in sensitivity during the 
forthcoming years:  a factor 6 better than the current upper limit for AMS and a
factor of 40 for GAPS.

Among the possible uncertainties mentioned in Barrau {\it et al.} 
(\cite{barrau4}), the most
important one was, by far, the possible existence of a QCD halo around 
{\sc pbh}s
(Heckler \cite{heckler}). The latest studies seem to show that this effect
should be much weaker (Mac Gibbon {\it et al.}, in preparation) than expected 
in Cline {\it et al.} (\cite{cline}). The results given in this
work should, therefore, be reliable from this point of view.

Nevertheless, two points
could make this picture a bit less exciting and deserve detailed 
studies. The first
one is related to secondary antideuterons: the cross-sections used 
in this work
could be slightly underestimated and some other processes could have 
to be taken into account
(Protassov {\it et al.}, in preparation).
This could increase the background  which should be considered with
the same propagation model. The
second one is that the signal is extremely close to the one obtained with the
annihilation of supersymmetric particles as the shape of the spectrum is mostly
due to fragmentation processes.
In the case of detection, it would be very
difficult to distinguish between the two possible phenomena, unless collider
data or 
indirect or direct neutralino dark matter searches
have given enough information to fix the supersymmetric parameters.\\

{\bf Acknowledgments}. We would like to thank K. Protassov
 and R. Duperray
for very interesting discussions about antideuteron cross-sections and 
C. Renault for her great help.

\end{document}